\documentclass[reprint,amsmath,amssymb,aps]{revtex4-2}

\usepackage{graphicx}
\usepackage{dcolumn}
\usepackage{bm}
\usepackage{xcolor}
\usepackage{mathcomp}

\begin{document}

\preprint{APS/123-QED}

\title{Insights on mass composition of UHECRs from small-scale anisotropy with large-scale structure simulations}

\date{\today}

\author{Ryo Higuchi}
\affiliation{Institute for Cosmic Ray Research, University of Tokyo, Kashiwa, Chiba 277-8582, Japan}
\affiliation{Astrophysical Big Bang Laboratory (ABBL), RIKEN Pioneering Research Institute (PRI), 2-1 Hirosawa, Wako, Saitama 351-0198, Japan}

\author{Eiji Kido}
\affiliation{Institute for Cosmic Ray Research, University of Tokyo, Kashiwa, Chiba 277-8582, Japan}
\affiliation{Key Laboratory of Modern Astronomy and Astrophysics, Nanjing University, Ministry of Education, Nanjing 210093, China}
\affiliation{Astrophysical Big Bang Laboratory (ABBL), RIKEN Pioneering Research Institute (PRI), 2-1 Hirosawa, Wako, Saitama 351-0198, Japan}

\author{Shigehiro Nagataki}
\affiliation{Astrophysical Big Bang Laboratory (ABBL), RIKEN Pioneering Research Institute (PRI), 2-1 Hirosawa, Wako, Saitama 351-0198, Japan}
\affiliation{RIKEN Center for Interdisciplinary Theoretical and Mathematical Sciences (iTHEMS), 2-1 Hirosawa, Wako, Saitama 351-0198, Japan}
\affiliation{RIKEN-Berkeley Center, RIKEN iTHEMS, University of California, Berkeley, Berkeley, CA 94720, USA}

\date{\today}

\begin{abstract}
Ultra-high-energy cosmic rays (UHECRs) with energies above 100 EeV provide important clues to the most energetic processes in the Universe.
At the highest energies, it is expected that UHECRs originate only from nearby sources because of energy losses during propagation.
In such a regime, small-scale anisotropies in arrival directions could appear if the magnetic deflections in cosmic magnetic fields are sufficiently small, for example, for a limited fraction of UHECRs with relatively low electric charge.
Using the Pierre Auger Observatory open data, we confirm that for events with energies above 100 EeV, there is no significant small-scale anisotropy within $3\tcdegree$, indicating an almost isotropic sky distribution.
To interpret this observation, we consider that heavier nuclei have higher photodisintegration thresholds due to their smaller Lorentz factors, resulting in longer propagation lengths. When combined with the nearly homogeneous distribution of galaxies at such distances, this can naturally explain the absence of anisotropies even above 100 EeV.
Based on this idea, we perform mock observations of UHECRs above 100 EeV assuming pure-proton, iron, and uranium compositions. We predict the expected UHECR distributions using the Millennium Run semi-analytic galaxy catalog and turbulent extragalactic magnetic fields.
By varying the source density $\rho_{\rm source}$ and magnetic field strength $\alpha$, we evaluate the anisotropy for each case.
Without magnetic fields, the observed isotropy above 100 EeV excludes the pure-proton and pure-iron cases.
The pure-uranium case can reproduce the observed isotropy when the source density is high and the magnetic field is turbulent.
When the turbulent field is strong, both the iron and uranium cases remain consistent with the data, whereas almost no parameter set can explain the observation for a pure-proton composition.
Our results constrain the possible combinations of source density and magnetic field strength across different UHECR compositions and provide a framework for interpreting future anisotropy measurements at the highest energies.
\end{abstract}

\maketitle
\section{Introduction} \label{sec:intro}
The origins of ultra-high-energy cosmic rays (UHECRs) remain a mystery.
Due to the energy loss of the UHECRs through their interaction with the cosmic microwave background (CMB) and infrared (IR) background, the origins of the UHECRs should be in the nearby universe \citep[GZK horizon]{1966PhRvL..16..748G,1966JETPL...4...78Z}.
Nevertheless, the origins of UHECRs have not yet been identified. 
The magnetic fields bend UHECR trajectories, creating a significant gap between the arrival directions of UHECRs on Earth and their sources, making it challenging to identify the sources.
The highest-energy cosmic rays among UHECRs were expected to be less susceptible to the effects of magnetic fields due to their small deflection angles. 
Previous theoretical studies on small-scale anisotropy have predicted that the highest-energy cosmic ray events would be concentrated around specific point sources, leading to the identification of the origins of UHECRs \citep{2003ApJ...586.1211Y,2003ApJ...592..311Y,2004ApJ...614...43Y, KACHELRIE2005486, Takami2006, Takami2008, Rouill2014}. 
Theoretical studies on small-scale anisotropies are believed to have insights into the source density of UHECRs. 
The Akeno Ground Air Shower Array (AGASA) reported a small-scale anisotropy of UHECRs above $4\times 10^{19}$ eV in \cite{Takeda1999}. They reported one triplet and three doublets within $2.5$ degrees. 
The AGASA experiment's report motivated studies of UHECR sources and next-generation UHECR observations over the past two decades. 

Due to the small flux, it is difficult to observe a UHECR event directly. 
The detection of extensive air showers induced by UHECRs enables us to observe them. 
Currently, two main experiments lead UHECR observation in the northern and southern hemispheres, respectively: the Telescope Array (TA) experiment \citep{2012NIMPA.689...87A} and the Pierre Auger Observatory \citep{Auger2015}. 
The Pierre Auger Observatory \citep{Auger2015} also searched for a small-scale excess of UHECRs above 60 EeV \citep{Abreu2013}. 
In \cite{Abreu2013}, they found no small-scale anisotropy and set the lower bounds of source density to be $\sim 10^{-4}\: {\rm Mpc^{-3}}$. 
In the Telescope Array (TA) experiment \citep{2012NIMPA.689...87A}, \cite{Abbasi2015} searched for a point-like source with observed data collected over five years above 0.5 EeV, and found no evidence for the source in the northern sky.
Although many of the highest-energy UHECRs have been detected, recent observations from the Pierre Auger Observatory and the TA experiment have not reported a clear point source, even for the highest-energy cosmic rays above 100 EeV \citep{Fujii:2024Ne}.
A single-event observation also supports this.
The highest energy UHECR in the TA experiment \citep[``Amaterasu'' event]{TA:2023} has been found to arrive from the direction of the ``void'' region in the nearby Universe. 
To explain these situations on arrival directions of UHECRs, we investigate the effects of source density, mass composition, and magnetic field on 100 EeV UHECR anisotropies. 
From the absence of a clear small-scale anisotropy, we estimate the lower limits of source density, mass composition, and magnetic field strength.

The ``heavier'' mass composition (usually assumed to be heavier than iron) is suggested by observations from both the Pierre Auger Observatory and the TA experiment \citep{Auger2025,2024PhRvD.110b2006A,2024PhRvL.133d1001A}. 
In particular, the observation of the slant depth of the air-shower maximum $X_{\rm max}$ constrains the mass number of primary cosmic rays. 
The Pierre Auger Observatory found the mass composition getting heavier above approximately 2 EeV \citep{Auger2025} from their $X_{\rm max}$ measurements. 
The TA experiment suggests a heavier component above 100 EeV, based on the arrival-direction distribution of UHECRs and assumptions about the large-scale structure and magnetic-field models \citep{2024PhRvL.133d1001A}. 
For the heavier composition case, the deflection by a magnetic field larger than the proton case causes a more isotropic distribution of UHECRs. 
In addition, one can expect a higher photodisintegration energy threshold due to the smaller Lorentz factors, which leads to a longer propagation length. 

This paper also considers the possibility that the highest energy cosmic rays are ``super heavy'', indicating nuclei heavier than $^{56}$Fe (iron). We took $^{238}$U (uranium) as the heaviest example of super-heavy nuclei that were detected in solar abundances.
These nuclei are generated in the r-process. 
The possibility of r-process nuclei in UHECRs has been suggested in previous work.
\cite{Metzger2011} suggested heavy nuclei from GRBs as UHECRs. 
Protomagnetars are also suggested to be r-process UHECR source candidates \citep{Bhattacharya2022, Ekanger2022}. 
Motivated by the kilonova observation \citep{2017Natur.551...80K} and ``Amaterasu'' event \citep{TA:2023}, the possibility of r-process UHECRs is under discussion in some reports \citep{Farrar2024, Zhang2026}. 
\cite{Farrar2024} and \cite{Zhang2026} suggest the possibility of super-heavy nuclei, and they are generated in the r-process from binary neutron star (BNS) mergers. 
\cite{Farrar2024} suggested the idea of BNS mergers as the UHECR source, and at the same time, \cite{Zhang2026} suggested the possibility that super-heavy UHECRs cannot be denied from a ``combined fit'' \citep{Aab_2017} with a current observational dataset of the Pierre Auger Observatory and TA experiment. 
Discussions of distant-source contributions are important for super-heavy UHECRs because they undergo longer propagation distances. 

In this study, we test ``light'', ``heavy'', and ``super-heavy'' (e.g., proton, iron, and uranium) UHECR scenarios using 100 EeV UHECRs from an independent method: self-consistent mock observations with N-body simulations.
We do not specify the source class for the UHECR origin and assume that the UHECR source distribution follows that of normal galaxies. 
Assuming an isotropic distribution of UHECRs at 100 EeV, we constrain the lower limits on source density, mass composition, and magnetic field strength. 
We briefly introduce observed arrival directions of UHECRs (Section \ref{sec:observation}).
In Section \ref{sec:analysis}, we introduce the self-consistent technique to generate mock UHECR datasets through a convolution of an N-body simulation catalog and cosmic-ray propagation.
We present the results of the mock observation and insights into source density and magnetic field strength in Section \ref{sec:results}.

\section{ANALYSIS}\label{sec:observation}
We adopt a number of ``doublets'' to evaluate the anisotropy on the arrival direction of UHECRs. 
We count the number of pairs $N_{\rm pair}$ in a dataset as some pairs within 3 $\deg$ radius :
\begin{equation}
    N_{\rm pair} = \frac{1}{2} \sum_{i=1}^{N_{\text{event}}} \left[ \left( \sum_{j=1}^{N_{\text{event}}} \delta \left( \lambda(\bf n, n_{j}) \leq \lambda_{0} \right) \right) \geq 2 \right]
\label{eq:Npair}
\end{equation}
and
\begin{eqnarray}
    \delta (\text{condition}) = \left\{
    \begin{array}{ll} 
    1 & \text{if condition is true} \\ 
    0 & \text{if condition is false,}
    \end{array}
    \right.
\end{eqnarray}
where $N_{\rm event}$, $\lambda$, $\lambda_0$ indicate the number of events in the dataset, angular distance between arrival directions $\bf n$, and the threshold of search radius (3 $\deg$), respectively. 
We define the radius $\lambda_0$ to be 3 degrees, following the Akeno Giant Air Shower Array (AGASA) studies \citep{AGASA1996} which reported doublets within 2.5 degrees separation. 
\begin{figure}
    \centering
    \includegraphics[width=0.45\textwidth]{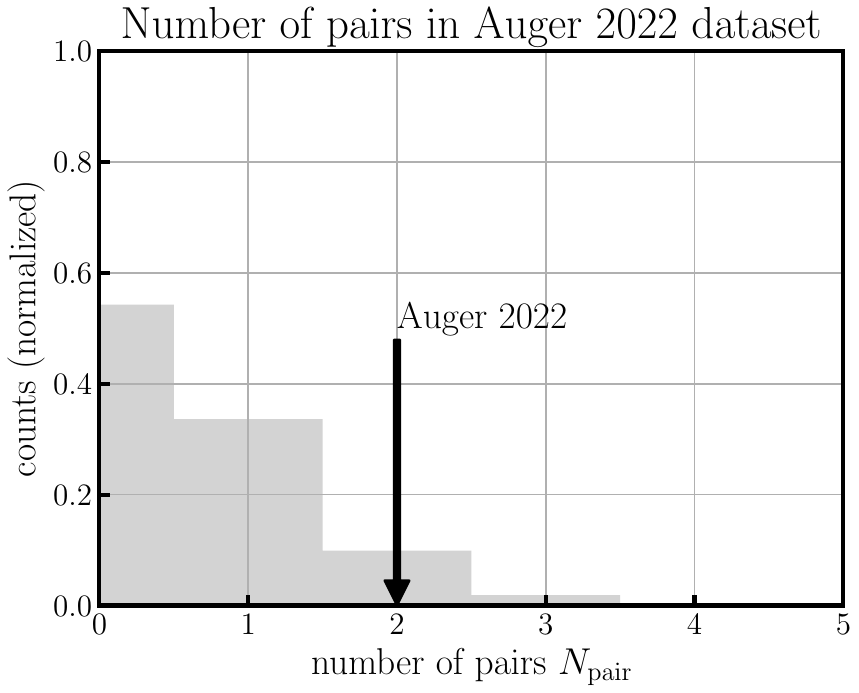}
    \caption{
    A pair count $N_{\rm pair}$ of the Auger Phase One dataset \cite{the_pierre_auger_collaboration_2021_4487613} (\textbf{black arrow}). 
    The grey histogram shows the number of pairs with an isotropic distribution. 
    }
    \label{fig:Auger}
\end{figure}
Figure \ref{fig:Auger} shows the number of pairs $N_{\rm pair}$ of the Auger Phase One dataset \cite{the_pierre_auger_collaboration_2021_4487613} (black arrow). 
We count the number of pairs $N_{\rm pair}$ for 35 observed UHECR events whose energies are more than 100 EeV.
For comparison with an isotropic distribution, we generate 100000 isotropic datasets at random, each with the same number of events as the observation. 
The number of pairs is $N_{\rm pair}=2.0$ (a black arrow in Figure \ref{fig:Auger}), which is located inside the isotropic $N_{\rm pair}$ distribution. 

\section{Simulation}\label{sec:analysis}
To interpret the observational result in Sec. \ref{sec:observation}, we conduct mock observations of UHECRs above 100 EeV assuming pure-proton, iron, and uranium compositions. 
\subsection{Energy loss distance of heavy UHECRs}\label{sec:eloss}
We calculate the energy-loss distances for uranium $^{238}$U as an example of a nucleus heavier than iron $^{56}$Fe, which is the heaviest species commonly considered in UHECR studies.
We use the cosmic-ray propagation code CRPropa~\cite{CRPropa3.2} for calculations.
Since CRPropa does not include information on $^{238}$U, we evaluate its energy-loss distances separately and compare them with those of protons and $^{56}$Fe obtained using CRPropa.
We calculate the energy loss due to photodisintegration by integrating photodisintegration cross sections computed with TALYS-2.0~\citep{2023EPJA...59..131K} over the photon spectra of the CMB and IRB~\citep{2012MNRAS.422.3189G}.
In the TALYS calculation, we consider photon energies from 0.2 to 200~MeV with a logarithmic spacing of $\Delta \log_{10}\varepsilon = 0.01$, and we enable the output of exclusive reaction channels with up to eight emitted particles.
The exclusive photodisintegration cross sections are post-processed to construct branching ratios.
We remove nonphysical channels (e.g., those with no emitted particles or invalid residual nuclei), and for each photon energy, the branching ratio of channel $i$ is defined as
\[
 b_i(\varepsilon) =
 \frac{\sigma_i(\varepsilon)}
      {\sum_j \sigma_j(\varepsilon)} \, .
\]
To reduce the number of channels, we retain only those whose branching ratio exceeds 5\% at least at one photon energy.
This procedure results in 21 photodisintegration channels for $^{238}$U, corresponding to the dominant emission channels of nucleons and light fragments.
The average mass loss per interaction is then evaluated as
\[
 \langle \Delta A \rangle =
 \sum_i b_i(\varepsilon)\,\Delta A_i \, ,
\]
where $\Delta A_i$ is the mass loss associated with channel $i$.
The total energy-loss rate is obtained by multiplying the interaction rate by the average fractional mass loss, $\langle \Delta A \rangle / A$.

\begin{figure}
      \centering
      \includegraphics[width=0.4\textwidth]{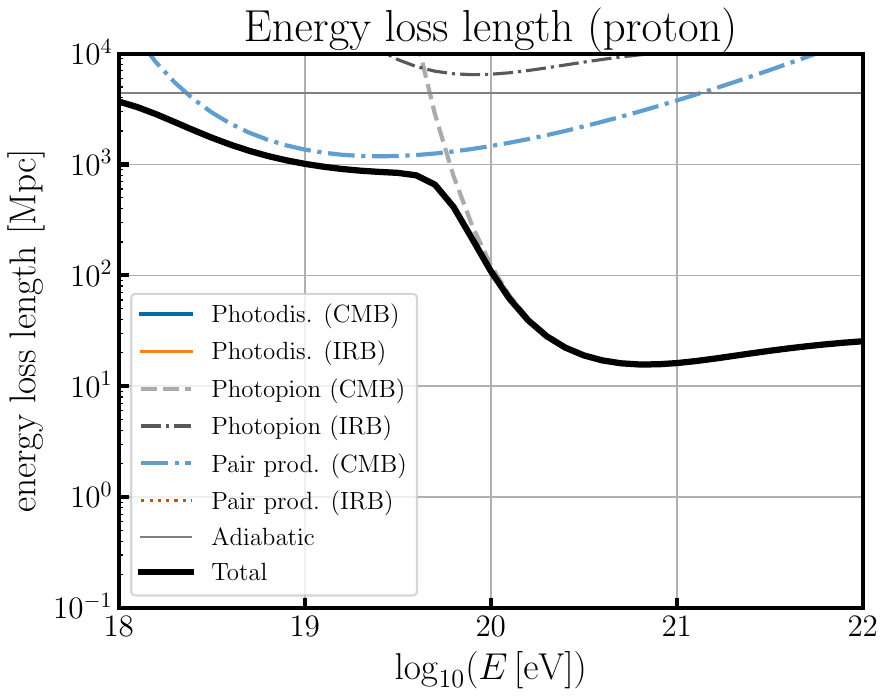}
      \includegraphics[width=0.4\textwidth]{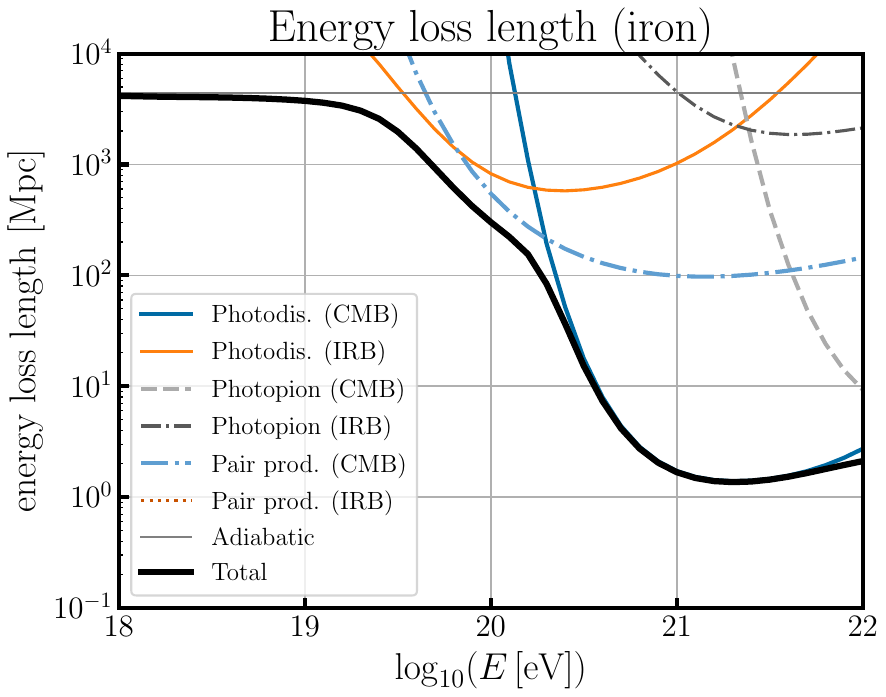}
      \includegraphics[width=0.4\textwidth]{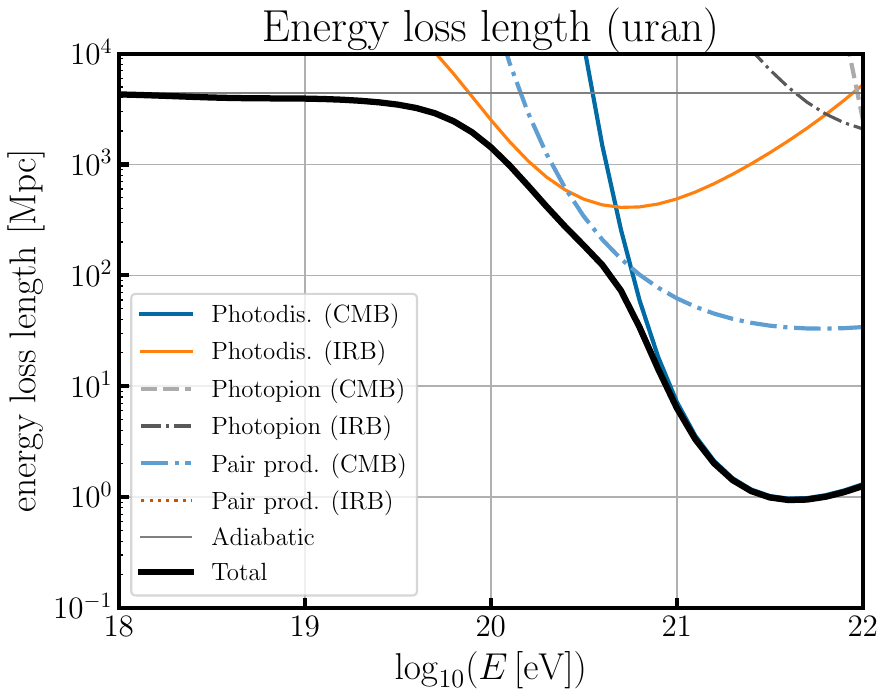}
      \caption{Energy loss distances of reactions of a proton (top panel), $^{56}$Fe (middle panel), and $^{238}$U (bottom panel) with CMB and IRB photons at redshift $z = 0$. 
      Solid black lines show the total energy-loss distances. 
      Solid gray lines indicate the adiabatic energy losses due to the cosmic expansion.
      }
      \label{fig:ElossLength} 
\end{figure}

Figure~\ref{fig:ElossLength} shows the energy loss distances of proton, $^{56}$Fe, and $^{238}$U.
We can see that the energy-loss distances for $^{56}$Fe are similar to those of $^{238}$U except for the energy scale.
The energies of background photons in the UHECR nuclei's rest frame are essential in the interactions, and energy loss distances are approximately rescaled by the mass number $A$ of UHECR nuclei. 
By comparing the energy-loss distances of $^{238}$U and $^{56}$Fe around 100 EeV, we obtained a rescaling factor of 2.6.

We performed one-dimensional propagation calculations of proton and $^{56}$Fe using CRPropa to estimate a distance $r$ distribution of sources $\rho (r)$ above 100 EeV. 
In the propagation, we assumed that the sources are uniformly distributed emitting cosmic rays with a power-law energy spectrum $dN/dE = E^{-2}$ for protons.
We assumed power-law energy spectra $dN/dE = E^{-1}$ for $^{56}$Fe and $^{238}$U.
We propagated 35,090/86,284/86,284 particles for pure-proton/iron/uranium cases.
The middle panel of Figure \ref{fig:sourcedistribution} shows the calculated source distribution $\rho (r)$ (as we explain in detail in Sec. \ref{sec:distribution}). 

\subsection{Distribution of matter (galaxies) in more distant universe} \label{sec:distribution}
\begin{figure*}
      \centering
      \includegraphics[width=0.4\textwidth]{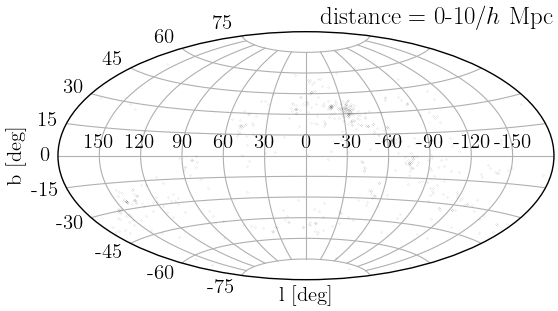}
      \includegraphics[width=0.4\textwidth]{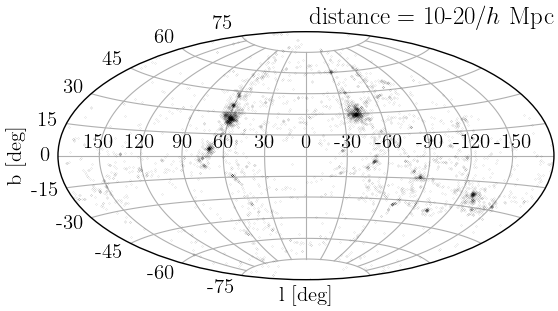}
      \includegraphics[width=0.4\textwidth]{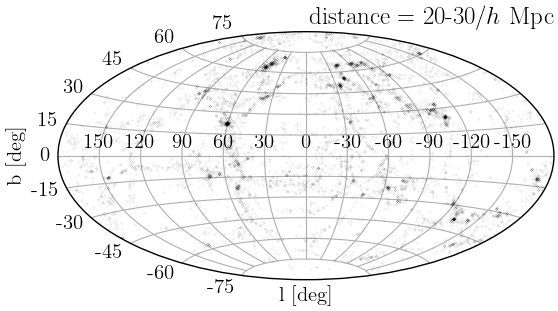}
      \includegraphics[width=0.4\textwidth]{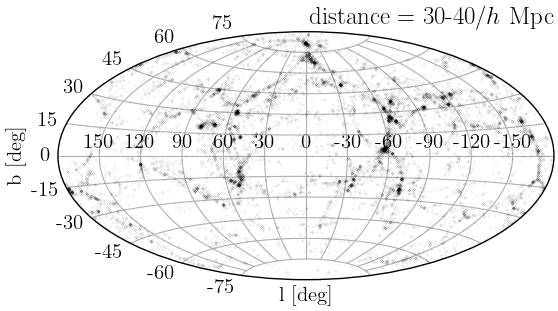}
      \includegraphics[width=0.4\textwidth]{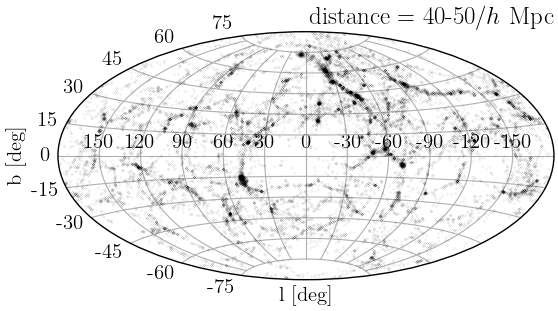}
      \includegraphics[width=0.4\textwidth]{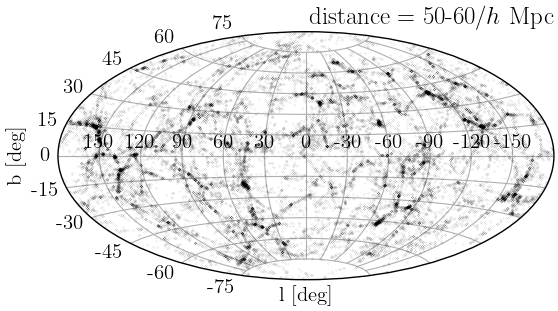}
      \includegraphics[width=0.4\textwidth]{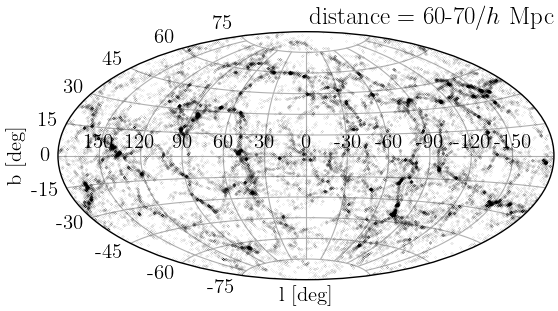}
      \includegraphics[width=0.4\textwidth]{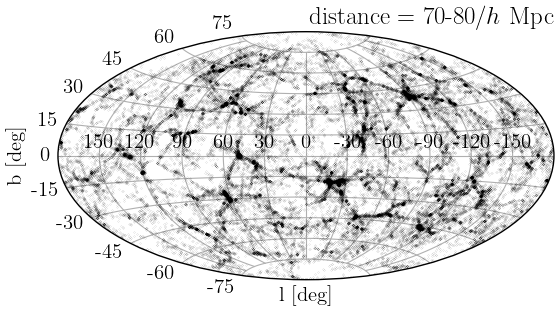}
      \includegraphics[width=0.4\textwidth]{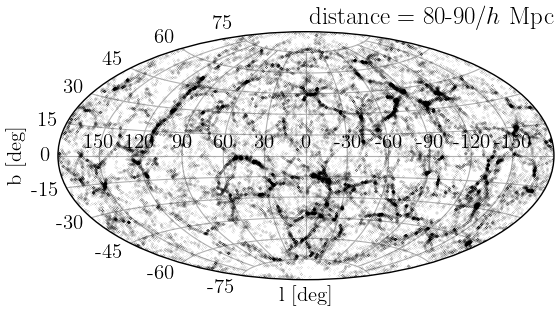}
      \includegraphics[width=0.4\textwidth]{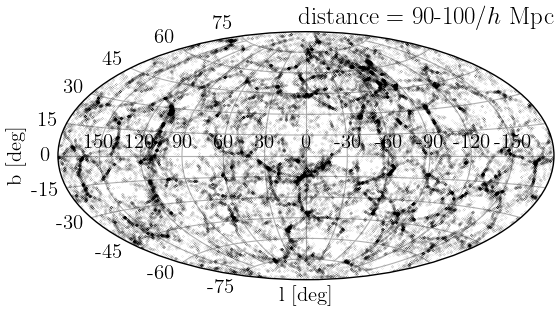}
      \caption{Simulated arrival directions from 0-10 Mpc/$h$ to 90-100 Mpc/$h$.
    Black points show the direction of semi-analytical galaxies in Millennium Run simulation \cite{2005Natur.435..629S}. 
    Because the number of semi-analytical galaxies at distance $r$ scales to $r^3$, we scale the color density of each panel so that the reader can distinguish the distribution.
      }
      \label{fig:anisotropy} 
\end{figure*}
\begin{figure}
    \centering
    \includegraphics[width=0.4\textwidth]{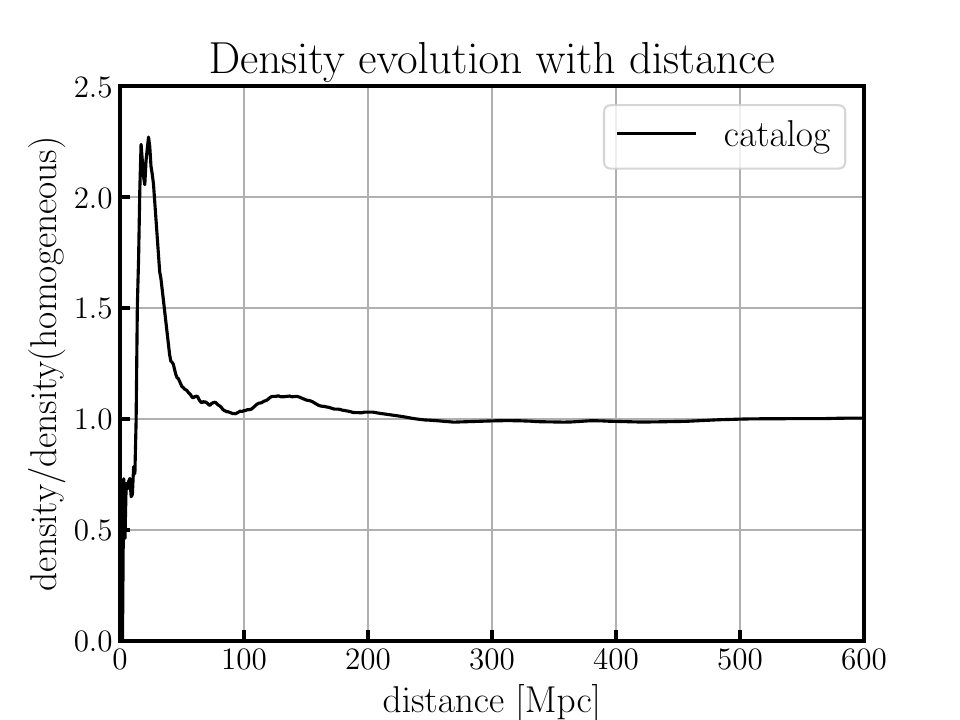}
    \includegraphics[width=0.4\textwidth]{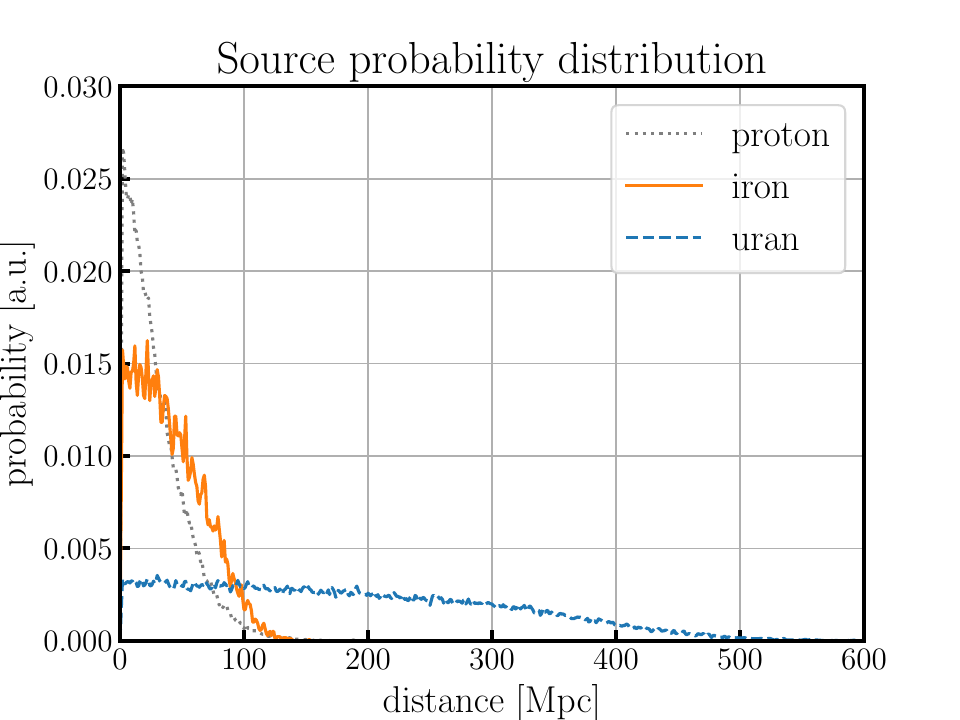}
    \includegraphics[width=0.4\textwidth]{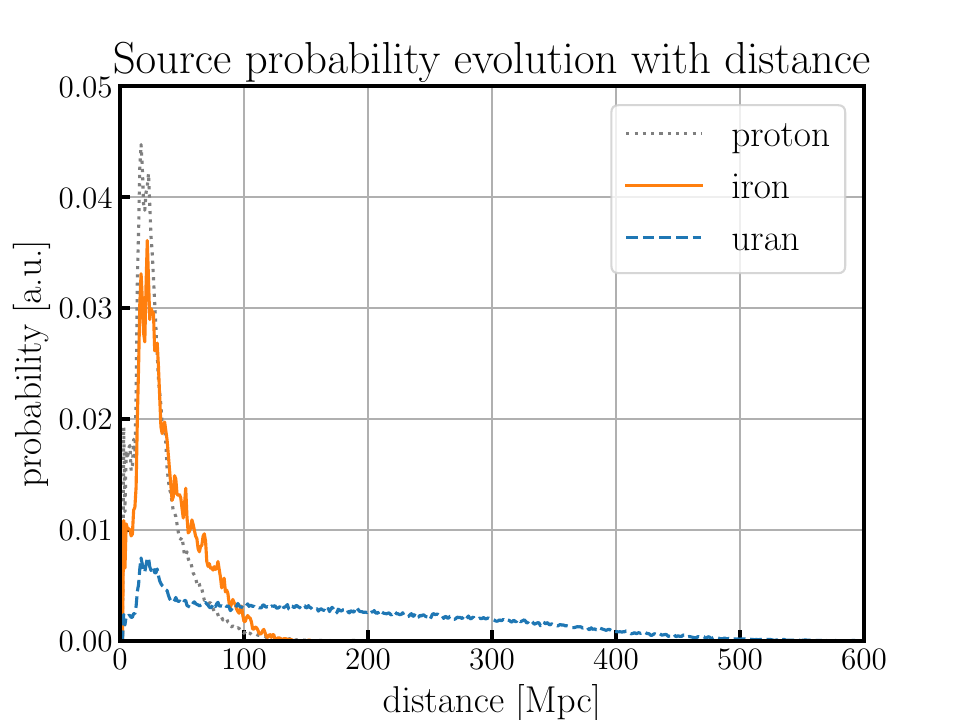}
    \caption{
    Matter density $m(r)$, calculated source distribution $\rho (r)$, and weighted source distribution $m(r) \times \rho (r)$ with a function of distance $r$.
    Gray-dotted, orange-solid, and blue-dashed lines show the probability distributions of sources for 100 EeV protons, iron, and uranium, respectively.
    The density is normalized to 1 when integrated over a distance of 1 Gpc.
    }
    \label{fig:sourcedistribution}
\end{figure}

To estimate the distribution of matter in a distant universe, 
we need to refer to a galaxy catalog that contains galaxies at distances of several hundred Mpc. 
Concerned about the completeness of distant galaxy detection, we do not adopt observed galaxy catalogs (e.g., 2MASS \citep{2MASS}).
Instead, we use a simulated galaxy catalog based on the Millennium Run semi-analytic model \citep{2005Natur.435..629S}.
The semi-analytic catalog is based on the $\Lambda$CDM simulation in a cube with a side length of $500 h^{-1}$ Mpc. 
We adopt $h=0.673$, same as in CRPropa 3.2 \citep{CRPropa3.2}. 
To enlarge the volume of the simulation, we connect $2^3$ boxes. 
We choose UHECR sources from the catalog with an assumption for source number density $\rho_{\rm source}$.
To convert the location of each source $(x,y,z)$ to the direction of the sky ${\bf n} = (l,b) $, we assume the observer is at eight vertices.%
The conversion of direction $(l,b)$ from $(x,y,z)$ is expressed as:
\begin{gather}
    r = \sqrt{x^2 + y^2 + z^2} \\
    \theta = \arccos\left(\frac{z}{r}\right)\\
    \phi = \text{sign}(y) \cdot \arccos\left(\frac{x}{\sqrt{x^2 + y^2}}\right)\\
    (l,b) = (\phi,\pi/2-\theta)
\end{gather}
The examples of the directional distribution ${\bf n} = (l,b)$ of semi-analytical galaxies are represented in Figure \ref{fig:anisotropy}.
We also show the matter density distribution $m(r)$ of the semi-analytical galaxies  with distance $r$ from the earth in the top panel of Figure \ref{fig:sourcedistribution}. 
We can see a bump in the matter density $m(r)$ in the top panel of Figure \ref{fig:sourcedistribution}, which corresponds to the clustering at distance $r=10-20/h$ Mpc, suggested in the top-right panel of Figure \ref{fig:anisotropy}.
As suggested in Figure \ref{fig:anisotropy}, the number density of matter becomes homogeneous beyond $\sim 200$ Mpc. 
The second panel in Figure \ref{fig:sourcedistribution} indicates the calculated source distribution of 100 EeV UHECRs $\rho (r)$ in Section \ref{sec:eloss}. 
Gray-dotted, orange-solid, and blue-dashed lines show the probability of source distribution for 100 EeV protons, irons, and uraniums, respectively.
We show the actual source distribution with distance $m(r) \times \rho (r)$ in the last panel in Figure \ref{fig:sourcedistribution}. 
In any mass composition case, the highest source density peak is in $r\sim 10-20/h$ Mpc, which reflects the highest peak of matter density $m(r)$. 
For both pure-proton and iron cases, the source distribution $m(r) \times \rho (r)$ is similar due to the relatively shorter propagation distances. However, the propagation distance for a pure-iron case is slightly further than that of a pure-proton case.
The source distribution in the pure-uranium case is different from the others. 
It extends to a greater distance, $r=500$ Mpc, due to the longer propagation distance. 
Note that the zero-point position $(x,y,z)=(0,0,0)$ in the simulation does not necessarily reflect the real Earth position. 
In this study, we prioritize establishing a theoretically self-consistent method for evaluating UHECR anisotropy using N-body simulations rather than a single realization. 
Determining Earth's position is an important question when discussing deflection by the coherent magnetic field structure, especially in the Galactic magnetic field. We discuss this topic further in a future study.
\begin{figure}
    \centering
    \includegraphics[width=0.4\textwidth]{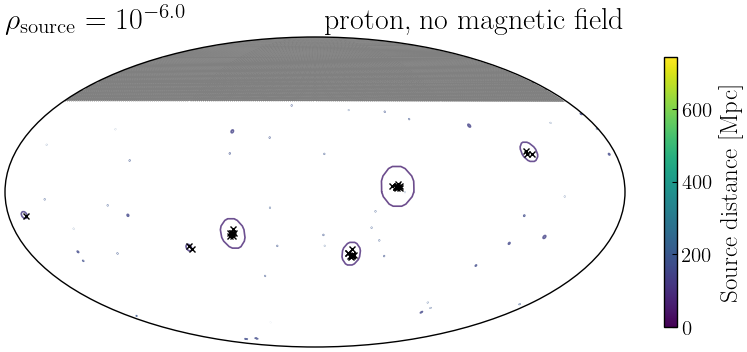}
    \includegraphics[width=0.4\textwidth]{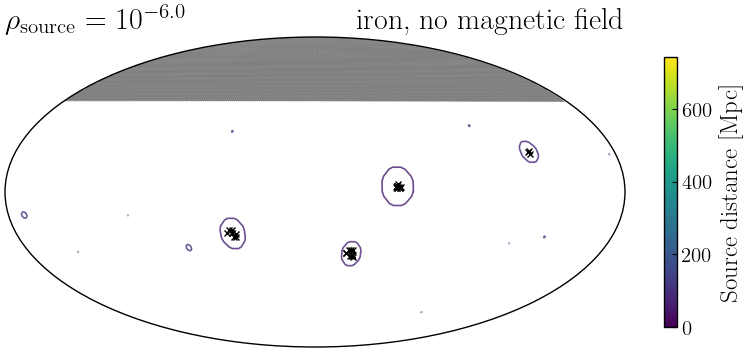}
    \includegraphics[width=0.4\textwidth]{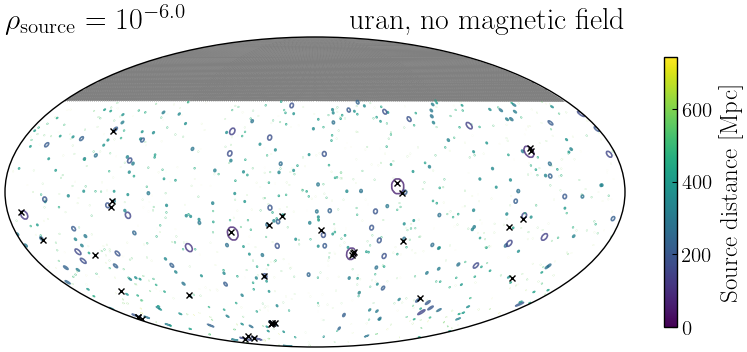}
    \caption{
    An example of a mock CR distribution without turbulent magnetic fields. 
    Colored circles indicate the direction of UHECR sources, which show high flux weight. 
    The area and color of each circle show the relative contribution and source distance, respectively. 
    Black crosses show the distribution of mock CR events. 
    We mask the area outside the observatory's FoV as a gray region. 
}
    \label{fig:dis_Moff}
\end{figure}

\subsection{Cosmic ray flux calculation}\label{sec:CRflux}
We calculate cosmic ray (CR) flux in the sky, weighting with the luminosity distance and energy loss distance $d_{\rm loss}$ calculated in the previous section (Figure \ref{fig:ElossLength}). The weight $w$ is written as
\begin{equation}
    w \propto r^{-2} \rho (r).
\end{equation}
To reflect the angular resolution of the detectors, we assume von-Mises Fisher (vMF) like distribution ${\rm vMF}({{\bf n}_{i},\theta_{\rm res}})$ with $\theta_{\rm res} = 1$ degree, centering the galaxy direction ${\bf n_{i}} = (l,b) $. 
We describe $
   {\rm vMF}({{\bf n}_{i},\theta_{\rm res}}) = \kappa  \exp (\kappa  \cos {\bf n_{\it i}} \cdot {\bf n}) / (4 \pi\sinh \kappa)$,
and approximate the $\kappa$ as $ \kappa = \frac{1}{2 (1-\cos(\theta_{\rm res}))}$ following \cite{Auger2022}.
We can calculate CR flux $F({\bf n})$ as a sum of flux from each source
\begin{equation}
    F({\bf n}) = {\sum_{i} w_i \times {\rm vMF}({{\bf n}_{i},\theta_{\rm res}})}, 
    \label{eq:CRflux2}
\end{equation}
and normalize 
\begin{equation}
       F({\bf n})^{\rm norm} =  \frac{F({\bf n})}{\sum_{\bf n}F({\bf n})}.
       \label{eq:CRflux}
\end{equation}
We generate CR events from the calculated CR flux $F({\bf n})$. 

We also consider the effects of a turbulent magnetic field. 
Following \cite{Waxman1996}, the smearing angle with a turbulent magnetic field $\theta_{\rm mag}$ is expressed as:
\begin{equation}
    \theta_{\rm mag} = 0.84 Z 
\left( \frac{r}{10\:\mathrm{Mpc}} \right)^{1/2} 
\left( \frac{\lambda}{1\:\mathrm{Mpc}} \right)^{1/2} 
\left( \frac{B}{1\:\mathrm{nG}} \right) 
\left( \frac{E}{100\:\mathrm{EeV}} \right)^{-1},
\end{equation}
same expression as in \cite{Bray_2018}. 
We change the parameter $\theta_{\rm res}$ to $\theta_{\rm mag}$ with a scaling of $Z  \left(\frac{r}{10\:\rm Mpc} \right)^{1/2} \left( \frac{\lambda}{1\:\rm Mpc} \right)^{1/2}  \left(\frac{B}{1\:\rm nG} \right) $ 
in Eq. \ref{eq:CRflux2}.
We define scale factor ${\rm \alpha} = \left( \frac{\lambda}{1\:\rm Mpc} \right)^{1/2}  \left(\frac{B}{1\:\rm nG} \right)$.
Based on the reference \citep{Bray_2018}, we decide the search range of the scale factor $10^{-6}$ to $\alpha=10$, which is comparable with the observational limits.
We consider three representative cases in figures  with $\alpha=1$ (strong), $10^{-3}$ (medium) and $10^{-6}$ (weak). 
Note that we adopt $\theta_{\rm mag}=\theta_{\rm res}=1$ degree when the smearing angle is smaller than the assumed angular resolution of detectors \citep{Auger2009}. 

\subsection{Mock CR datasets generation}\label{sec:CRdataset}
From the assumption of the CR flux map in Eqs. \ref{eq:CRflux2}-\ref{eq:CRflux}, 
we generate mock UHECR events for each dataset.
We repeat the same process and generate 1000 datasets for each assumption. 
Following Eq. \ref{eq:Npair}, we count the number of pairs in each dataset and investigate the distribution. 
For comparison, we also generate isotropic events in the same manner as in Sec. \ref{sec:observation} and check the distribution of $N_{\rm pair}$ of 1000 datasets.  
We also test with different source number density $\rho_{\rm source}$ case: $\rho_{\rm source} = 10^{-6}, 10^{-5.5}, 10^{-5}, 10^{-4.5}, 10^{-4}, 10^{-3.5}, 10^{-3}, 10^{-2.5} \: {\rm Mpc}^{-3}$ case. This range of source density $\rho_{\rm source}$ covers the density of Milky-Way-like galaxies, AGNs, starburst galaxies \citep{Bister2024}. 

For a fair comparison with the result of the observational dataset in Sec. \ref{sec:observation}, we apply the field of view (FoV) of the Pierre Auger Observatory. 
We follow equations in \cite{Sommers2001} to calculate the FoV:
\begin{eqnarray}
\label{eq:sommer}
\omega(\delta) \propto {\rm cos}(a_0){\rm cos}(\delta){\rm sin}(\alpha_{\rm m})+\alpha_{\rm m} {\rm sin}(a_0){\rm sin}(\delta), 
\end{eqnarray}
and
\begin{align}
\alpha_{\rm m}=
\begin{cases}
    0\: & (\xi>1) \\
    \pi\: & (\xi<-1) \\
    {\rm cos}^{-1}(\xi)\: & (-1<\xi<1)
\end{cases}
\end{align}
\begin{eqnarray}
\xi=\frac{ {\rm cos}(\theta_{\rm m})-{\rm sin}(a_0){\rm sin}(\delta)}{{\rm cos}(a_0){\rm cos(\delta)}}, 
\end{eqnarray}
where $\theta_{\rm m}$ is the maximum zenith angle and $a_0$ is the latitude of the observatory. 
We adopt $\theta_{\rm m}=80\tcdegree$ and $a_0=-35.2\tcdegree$ to produce mock datasets for the Pierre Auger Observatory. 
Note that we treat declination $\delta$ and galactic latitude $b$ as the same parameter for simplicity. 
 We generate 35 events over the FoV of Pierre Auger Observatory for each dataset, which is the same number of events as in the observational dataset in Sec. \ref{sec:observation}.

Figure~\ref{fig:dis_Moff} shows an example of a mock CR distribution without turbulent magnetic fields. 
In the pure-proton and iron cases, nearby sources make a major contribution to the observed UHECRs. 
Due to the absence of deflection by magnetic fields, events concentrate on their sources. 
In the pure-uranium case, the source and event distribution becomes more isotropic due to the contribution of distant sources. 

\begin{figure}
    \centering
    \includegraphics[width=0.4\textwidth]{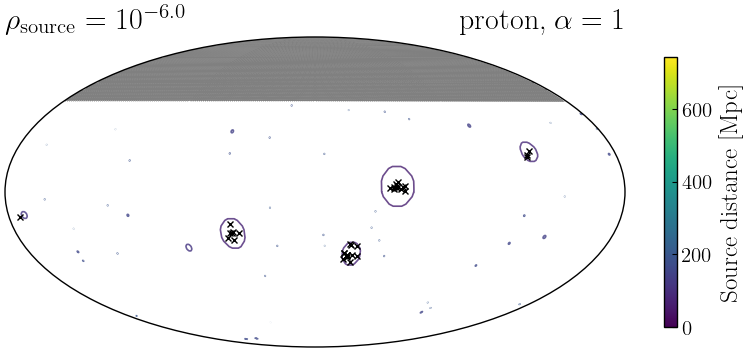}
    \includegraphics[width=0.4\textwidth]{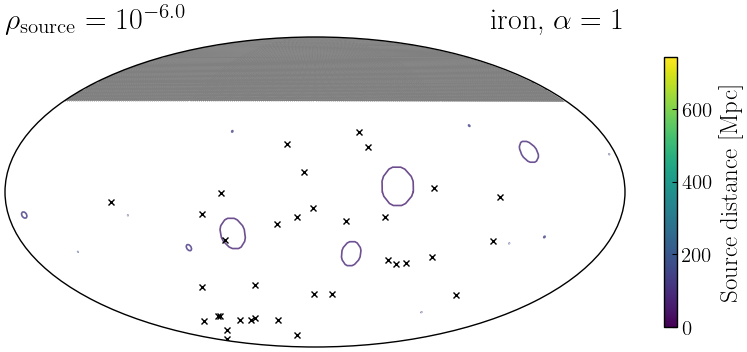}
    \includegraphics[width=0.4\textwidth]{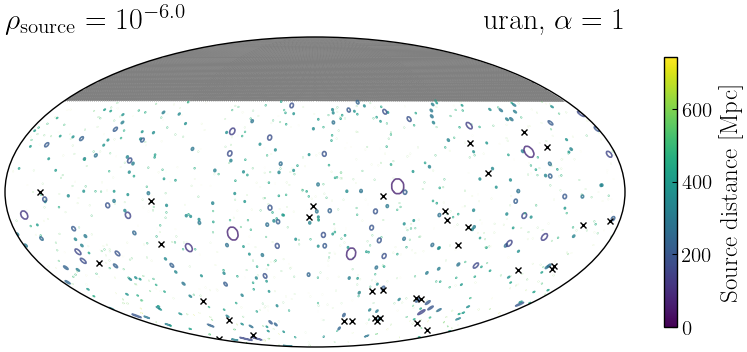}
    \caption{
    Same as Figure \ref{fig:dis_Moff}, but in cases with turbulent magnetic fields.
    }
    \label{fig:dis_Mon}
\end{figure}

Figure \ref{fig:dis_Mon} shows the example of a mock distribution with turbulent magnetic fields ($\alpha=1$) from the same sources in Figure \ref{fig:dis_Moff}.
Due to the shorter distance from the sources and smaller deflection angles, the CR events concentrate around their sources in the pure-proton case. 
In the pure-iron case, however, the distribution becomes more isotropic due to larger deflection angles. 
The CR distribution in the pure-uranium case favors isotropy because of contributions from distant sources and large deflection by magnetic fields.

\section{results}\label{sec:results}
We test the anisotropy of mock UHECR events in the same manner as the observational dataset with Eq. \ref{eq:Npair}.
\begin{figure*}
    \centering
     \includegraphics[width=\textwidth]
     {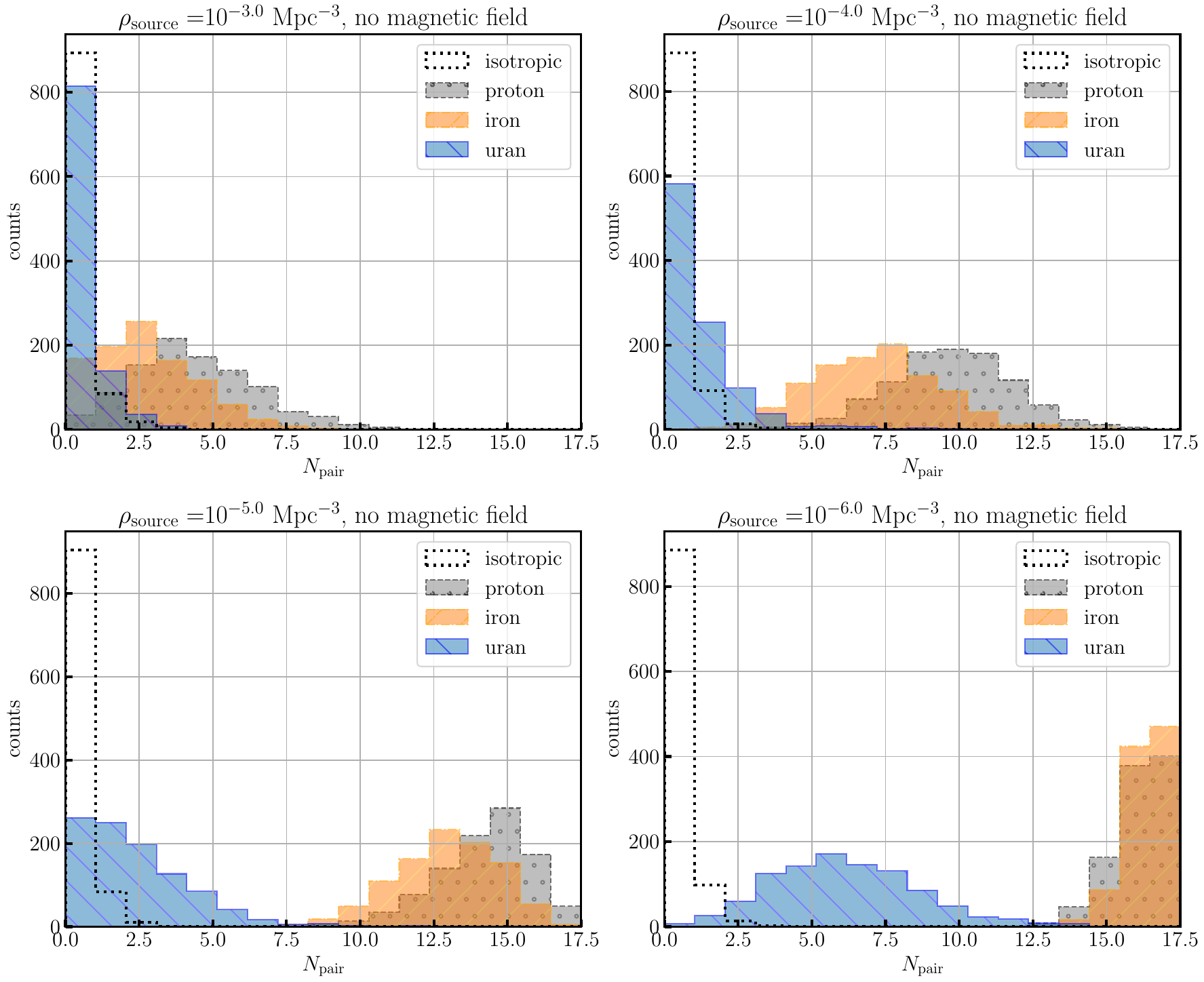}
    \caption{
    Histograms of the number of pairs $N_{\rm pair}$ for MC datasets from Millennium Simulation. 
    From the top to bottom, we change the source number density from $10^{-6}$ to $10^{-3}$ per $\rm Mpc^3$, respectively. 
    The gray, orange, and blue histograms show the distributions of pure-proton, pure-iron, and pure-uranium cases, respectively. 
    The black-dotted outline indicate the isotropic distribution.
    }
    \label{fig:hist_MFoff}
\end{figure*}
First, we show the result without the magnetic field assumption, in which UHECR arrival distribution reflects the galaxy distribution without magnetic deflections. 
Figure \ref{fig:hist_MFoff} shows the distribution of $N_{\rm pair}$ for 1000 datasets without the magnetic field assumption. 
The pure-proton case cannot reproduce the isotropic distribution for any source density considered in this figure. 
The pure-iron case shows a similar trend to the pure-proton case.
Only a pure-uranium case reproduces the isotropic distribution when source density is high (ex. $\rho_{\rm source}=10^{-3.0}$ and $10^{-4.0}$ cases).

\begin{figure*}
    \centering
    \includegraphics[width=\textwidth]{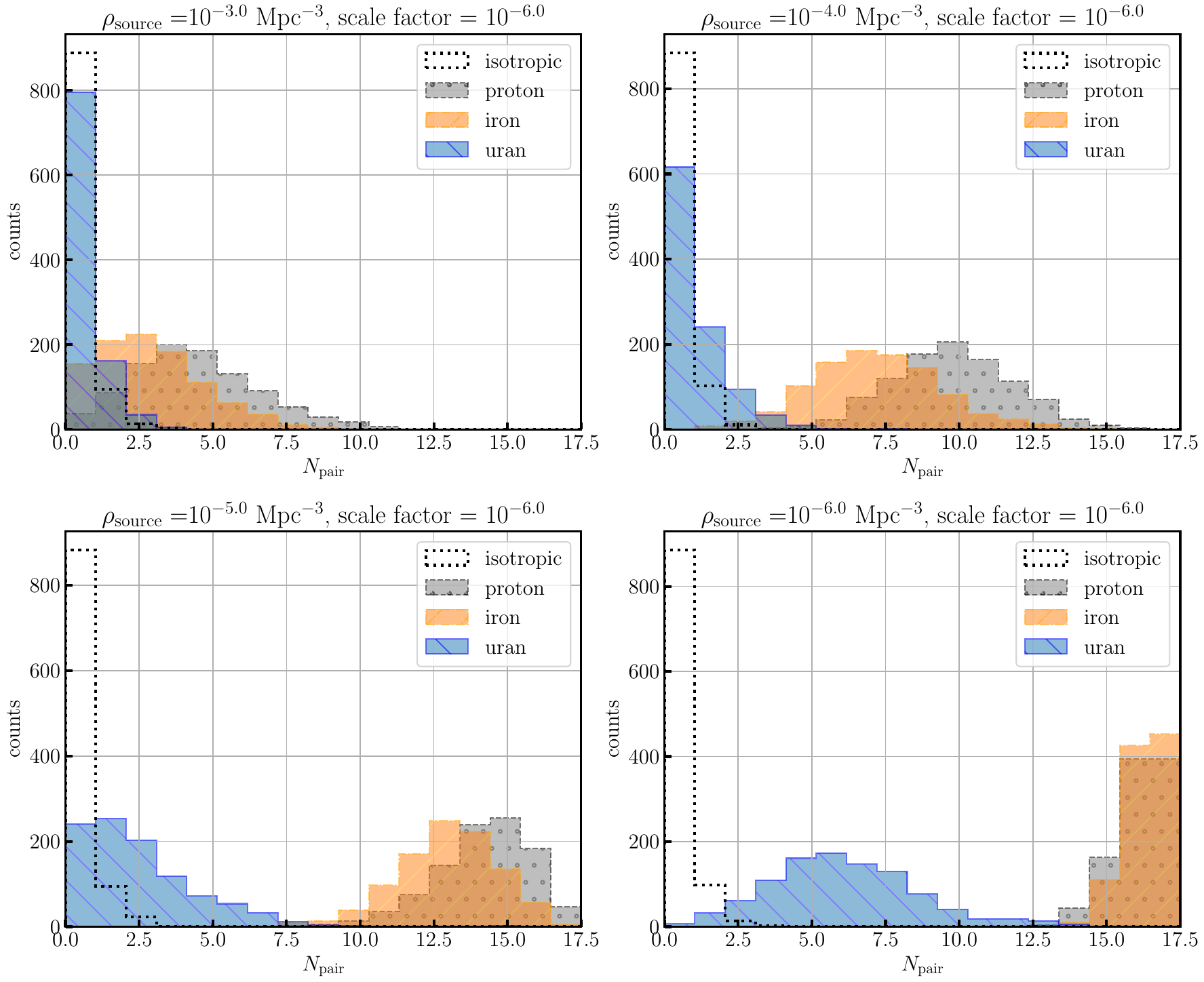}
    \caption{
    Same as Figure \ref{fig:hist_MFoff}, but for the case when we assume turbulent magnetic fields (weak MF).
    }
    \label{fig:hist_MFon}
\end{figure*}

\begin{figure*}
    \centering
    \includegraphics[width=\textwidth]{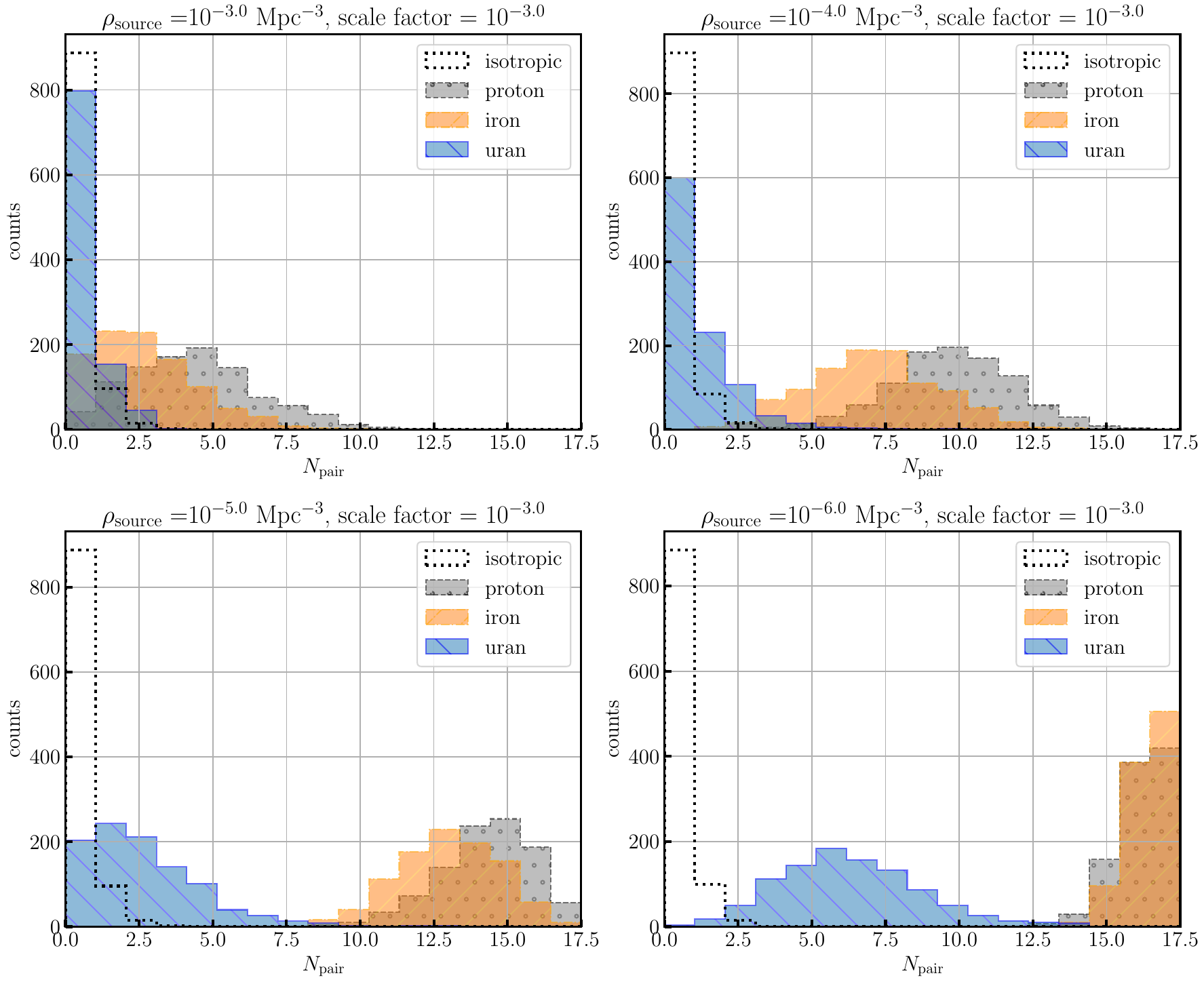}
    \caption{
    Same as Figure \ref{fig:hist_MFon}, but for the case when we change the magnetic field strength (medium MF). 
    }
    \label{fig:hist_MFon2}
\end{figure*}

\begin{figure*}
    \centering
    \includegraphics[width=\textwidth]{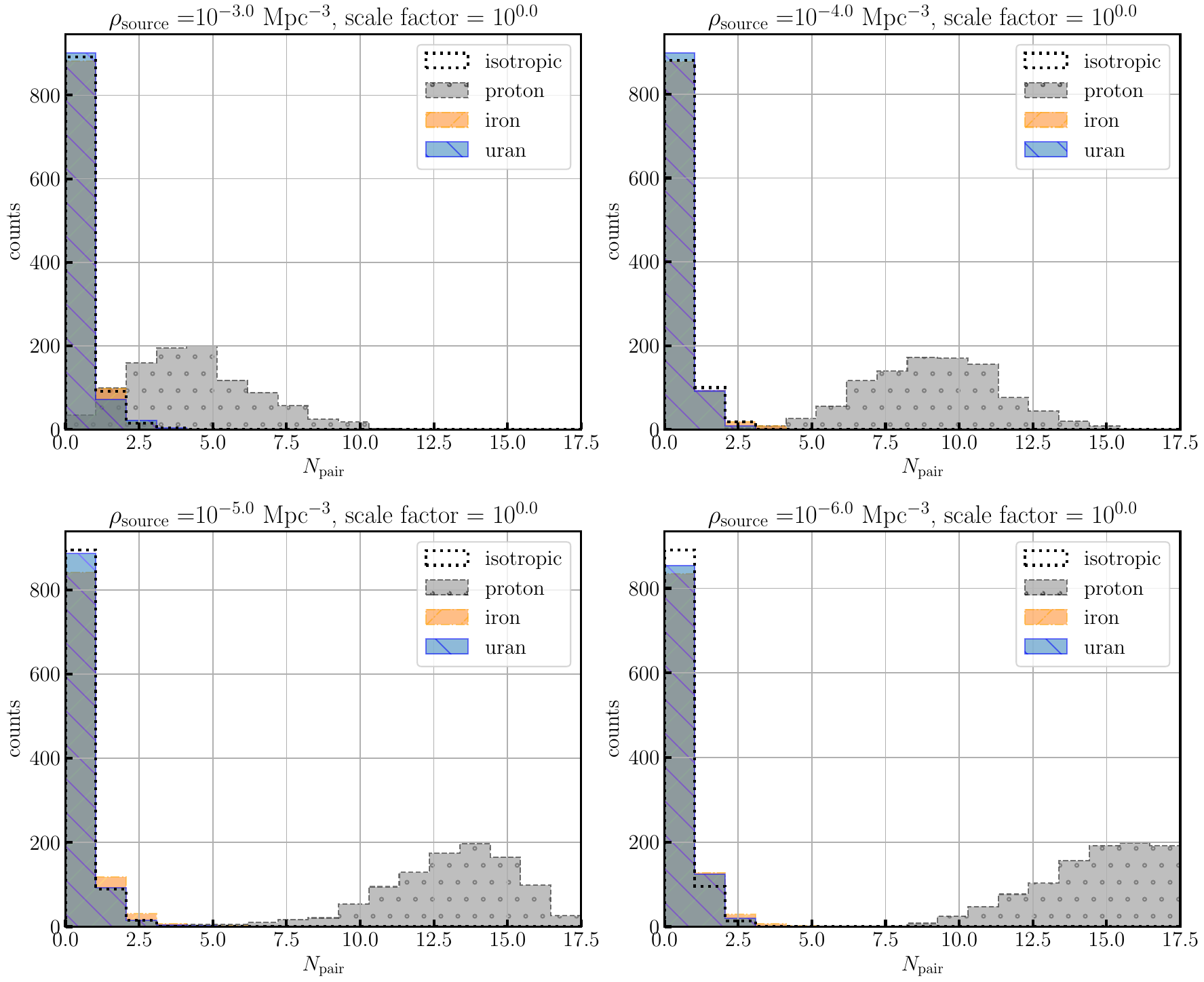}
    \caption{
    Same as Figure \ref{fig:hist_MFon}, but for the case when we change the different turbulent magnetic field scale (strong MF).
    }
    \label{fig:hist_MFon3}
\end{figure*}

We also calculate $N_{\rm pair}$ and test the anisotropy with an assumption of a turbulent magnetic field. 
Figures \ref{fig:hist_MFon}-\ref{fig:hist_MFon3} show the distribution of $N_{\rm pair}$ for 1000 datasets with turbulent magnetic fields. 
We show the strong magnetic field (MF) case, medium MF case, and weak MF case in Figure \ref{fig:hist_MFon}, \ref{fig:hist_MFon2}, and \ref{fig:hist_MFon3}, respectively.
In the pure-proton case, there is still a discrepancy between the isotropic distribution and the CR distribution. 
On the other hand, the pure-iron and pure-uranium cases show isotropic distributions due to scattering by magnetic fields (scale factor ${\rm \alpha} = 1$). 
When we assume smaller effects by turbulent magnetic fields (${\rm \alpha} = 10^{-3}$--$10^{-6} $), even a pure-uranium case cannot reproduce an isotropic distribution. 
We note that the pure-iron/uranium case shows isotropic distribution (when scale factor ${\rm \alpha} =1$) for different reasons. 
In the pure-iron case, the dominant sources are nearby galaxies, as suggested in Figure \ref{fig:sourcedistribution}. 
Due to the large scattering angle by turbulent magnetic fields, UHECRs from the nearby source become isotropic. 
In the pure-uranium case, distant sources up to 500 Mpc away can contribute to the observed arrival direction. 
In addition to the scattering angle by the turbulent magnetic fields, isotropic distant sources can contribute to the isotropy. 

\begin{figure*}
    \centering
    \includegraphics[width=0.8\linewidth]{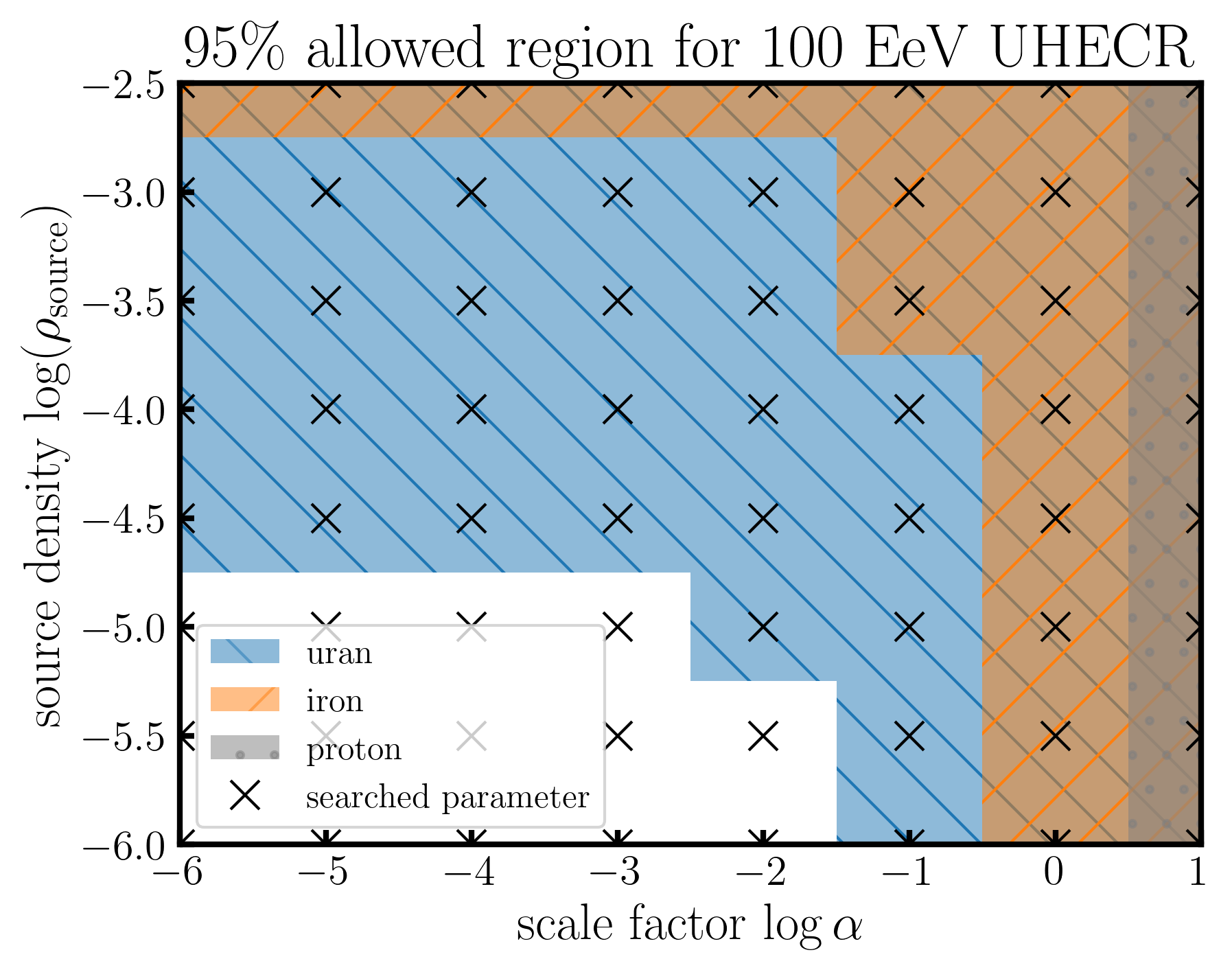}
    \caption{
    Allowed parameter space for source density and scale factor $(\alpha,\rho_{\rm source})$. 
    Black crosses indicate the parameter points we search. 
    Grey, orange, and blue regions show the allowed parameter space $(\alpha,\rho_{\rm source})$.
    } 
    \label{fig:space}
\end{figure*}

With no detection of a significant doublet in simulations, we can place lower limits on the source density and magnetic field strength.
We define the ``allowed region'' using the $N_{\rm pair}$ distribution of isotropic (isotropic mock dataset) and anisotropic (mock dataset with proton/iron/uranium) samples.  
If the median value of $N_{\rm pair}$ with an anisotropic sample is inside the one-sided 95$\%$ of the isotropic distribution, we define the parameter set $(\alpha,\rho_{\rm source})$ as an allowed region with $95\%$ C.L.
Figure \ref{fig:space} shows the allowed parameter space of $(\alpha,\rho_{\rm source})$, which can produce the isotropy. 
Almost all the parameter sets $(\alpha,\rho_{\rm source})$ reject the pure-proton case. 
The pure-uranium case can explain the current observation with turbulent magnetic fields when the source density is high ($\rho=10^{-4.5}\: {\rm Mpc}^{-3}$). When the turbulent field is strong, the pure-iron and uranium cases are consistent with the observations. 

\section{Discussions} 
\subsection{Rejection of pure-proton case}
As shown in Figure \ref{fig:space}, the pure-proton case is allowed only when the turbulent magnetic field is extremely strong ($\alpha =10$).  
It is well explained that the magnetic fields must be strong enough to erase the anisotropies that would otherwise arise from galaxy clusters within the GZK horizon.
The rejection of the pure-proton case is consistent with current observational trends with $X_{\rm max}$ measurements by the Pierre Auger Observatory \citep{Auger2025}, which indicate a heavier mass composition at higher energies.
Current observational limits on GZK photons and neutrinos provide the strongest existing constraints on the proton component of UHECRs, although they do not yet exclude a pure-proton composition above 100 EeV.
The non-detection of GZK photons by the Pierre Auger Observatory and the Telescope Array places the tightest limits around 100 EeV, where a pure-proton composition would otherwise produce an observable photon flux through the GZK process. 
Similarly, the upper limits on GZK neutrinos from Auger and TA provide complementary constraints on the proton fraction in this energy range. 
Finally, recent results from IceCube \citep{IceCube2025}, which set the most stringent upper limits on the proton fraction above 30 EeV using high-energy neutrino observations, are consistent with these results.
Future studies with higher statistics on higher-energy neutrinos are expected to set an upper limit on the proton fraction in the higher-energy range focused on in this paper. 

\subsection{Possibility of super-heavy UHECRs}
In this paper, we propose that the highest-energy cosmic rays above 100 EeV are composed of super-heavy nuclei and originate from a more distant universe than previously thought, making the arrival-direction distribution more isotropic and consistent with current observations. 
Our hypothesis in this study will be tested by future observations of the highest-energy cosmic rays and analysis of their chemical composition. We also note that the chemical composition is altered by interactions between the highest-energy cosmic rays and the CMB/IR. 
The super-heavy UHECRs will interact with background photons, producing secondary gamma rays and neutrinos during propagation.
Future observations of secondary gamma rays and neutrino fluxes will also play important roles in constraining the chemical composition. 

A simple count of doublet or small-scale anisotropy cannot distinguish the mass composition of UHECRs above 100 EeV when the turbulent magnetic fields are strong. 
Large-scale anisotropy studies above 100 EeV may provide information on the source population and on scattering by turbulent magnetic fields \citep{Urban2021, Allard2022}, which should be our next topic. 

\subsection{Discussions on source populations}
It is predicted that small-scale anisotropy studies provide an insight into source populations \citep{Takami2006, Takami2008}.
\cite{Takami2006} predicted the most appropriate source density to be $\rho_{\rm source}=5\times 10^{-6}\: \rm Mpc ^{-3}$, based on the reports of small-scale anisotropies in the AGASA observation above $10$ EeV \citep{Takeda1999}. 
\cite{Takami2008} also investigated the reproducibility of isotropy of UHECRs, and suggested the possibility of number density to be $\rho_{\rm source}=10^{-2}$--$10^{-3} \: \rm Mpc ^{-3}$. In these studies, they assume pure proton composition. 
\cite{Abreu2013} reported the lower limit of source density to be $\sim 10^{-4}\:\rm Mpc^{-3}$ from the arrival directions of UHECRs.

Although the assumed energy threshold and mass composition differ, previous studies do not contradict our results. 
Another recent work has reported constraints on source density from a different approach to small-scale anisotropy and is consistent with our results. 
\cite{Bister2024} constrained the source density and EGMF from the dipole amplitude obtained with the Pierre Auger Observatory. 
\cite{Kuznetsov2024} derived the lower bound to be $1.0 \times 10^{-4} \: \rm Mpc^{-3}$ from a large-scale structure model and the highest event in the TA experiment \citep{TA:2023}. 
\cite{2024PhRvD.110b2006A} constrained source density to be $>2\times10^{-5}\: \rm Mpc^{-3}$ from the observed arrival directions of UHECRs above 100 EeV with the TA experiment. 
Further comparisons are necessary in future work.


\subsection{Realization of simulations}\label{sec:realization}
\begin{figure*}
    \centering
    \includegraphics[width=0.4\linewidth]{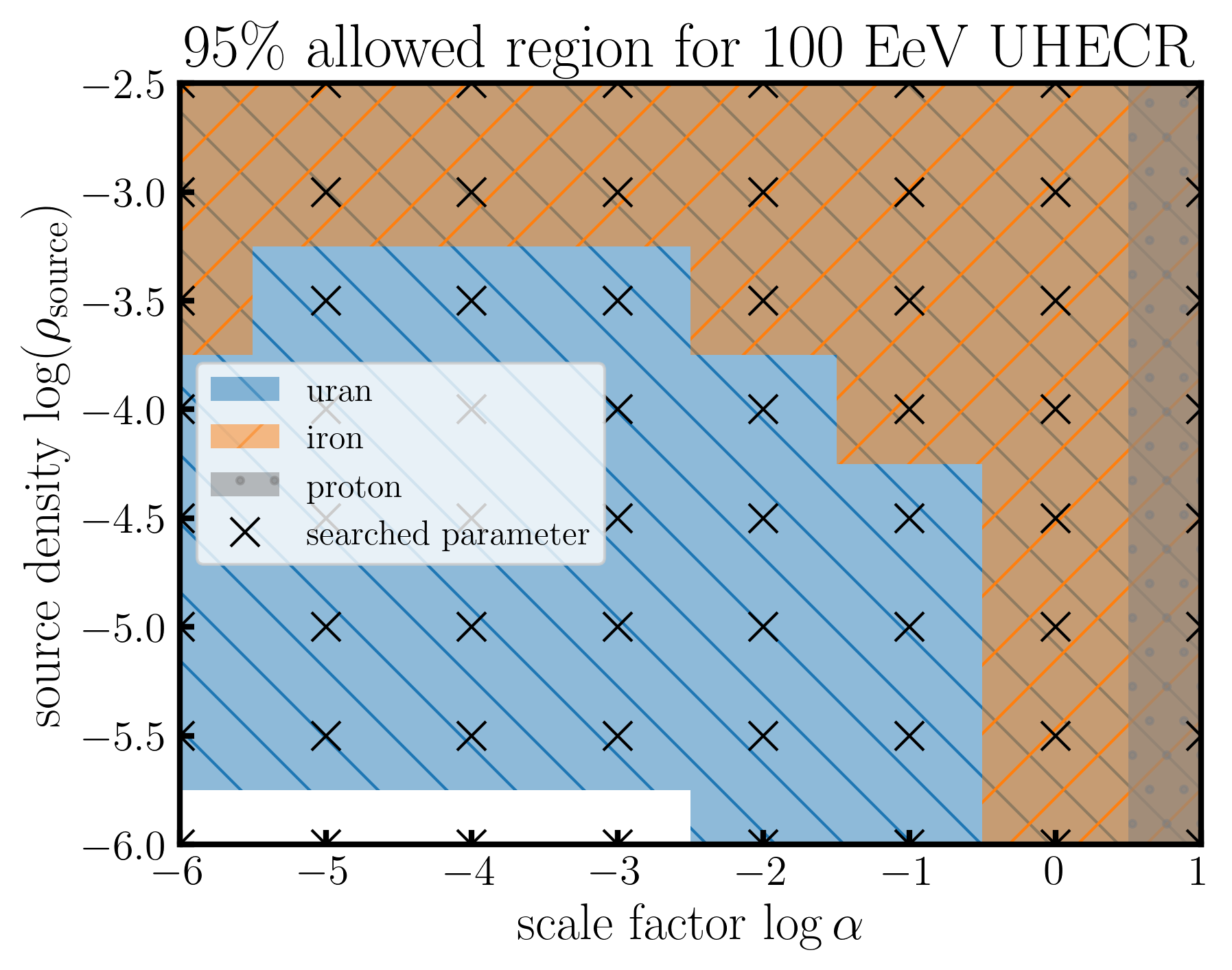}
    \includegraphics[width=0.4\linewidth]{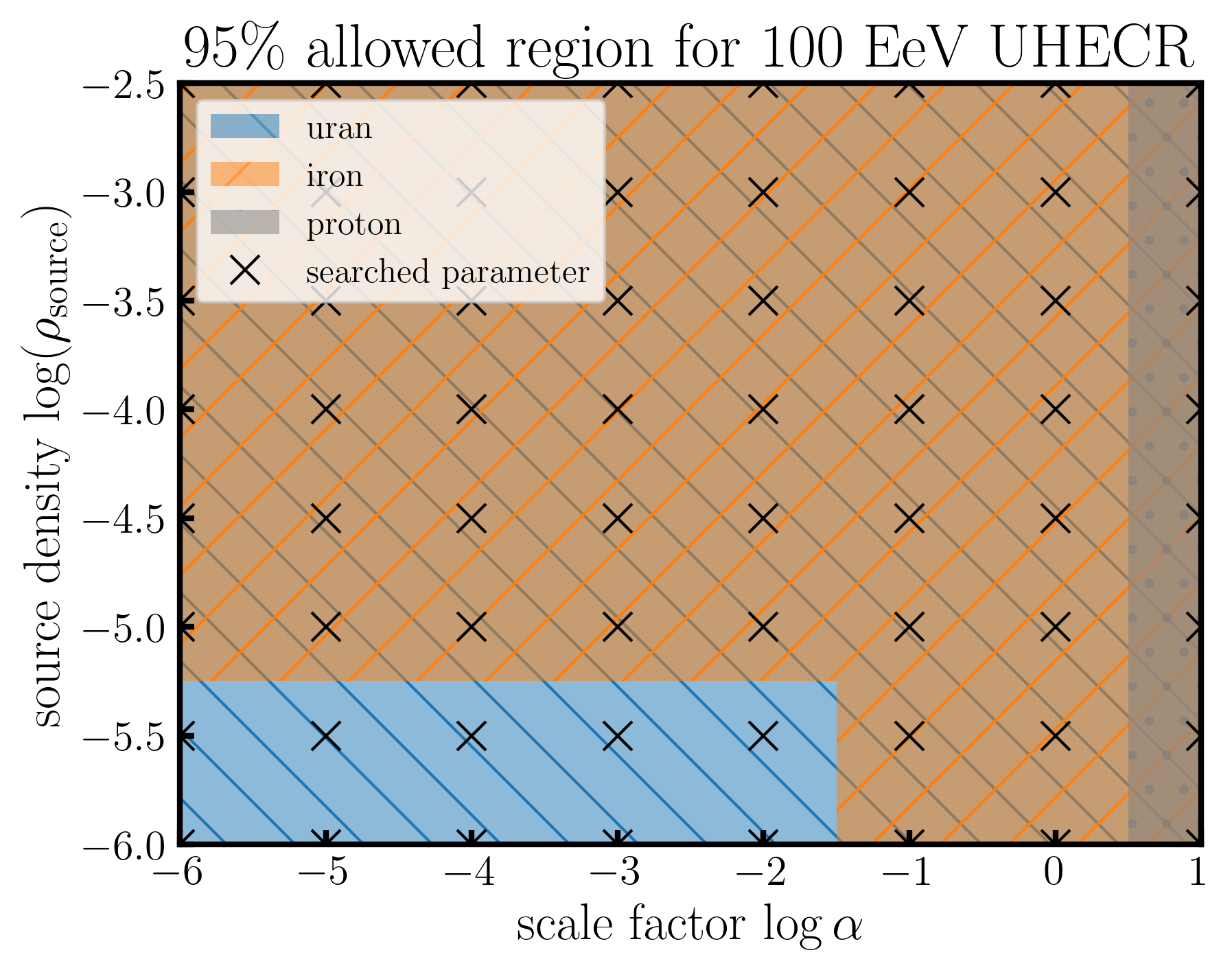}

    \caption{
   Same as \ref{fig:space}, but with JF12 model \citep{JF12a, JF12b}.
   Left and right panels show without and with turbulent fields, respectively. 
    } 
    \label{fig:space_JF12}
\end{figure*}
Although we focus on establishing a simple self-consistent method to evaluate the reproducibility of isotropy, the realization of simulations is an important issue relative to observations.

In this study, we assume a simple turbulent magnetic field in \cite{Waxman1996, Bray_2018}. 
It is suggested that magnetic fields have structure, and discussions on anisotropy with Galactic magnetic field (GMF) models are mandatory \citep{JF12a, JF12b}. 
It is suggested that coherent deflection by the magnetic fields affects the anisotropy analysis \citep{Higuchi2023}.

We regenerate the mock UHECR datasets from the same galaxy catalog in Sec. \ref{sec:analysis}. 
We apply the ``GMF lens'' in CRPropa 3.2 \citep{CRPropa3.2} to the CR flux, which is calculated in Sec. \ref{sec:CRflux}. 
We adapt the JF12 model \citep{JF12a, JF12b} as the GMF model. 
We show examples with the JF12 model in Figure \ref{fig:space_JF12}. 
The regular GMF model decreases the lower limit of source density by one magnitude. 
The turbulent field allows for a much wider parameter space than the regular GMF model. 
Recently, there have been reported updates to GMF models \citep{UF2024, KST2024}; large systematic differences remain among the models. 
Further comparison of the effects of magnetic fields, using the updated GMF models, is necessary.

The source distribution and extragalactic magnetic field structure along a large-scale structure are other questions for realization. 
We do not adapt the observed galaxy catalog (e.g., 2MASS \cite{2MASS}) to account for the detection completeness of distant galaxies. The appropriate reproduction of the actual local Universe is one challenging problem. 
We need to note that the result in this paper is only one set of realization: a position of the observer and the coordinate system. 
There is a degree of freedom in how one choose the position of the observer within the cube with a side length of $500h^{-1}$ Mpc. Even if we choose the one point of observer, there is still flexibility to define the coordinate system. The flexibility of coordinate system affects the position of the FoV of the observatory and the GMF models. 
The modeling of extragalactic magnetic fields is also a challenging question. 
The extragalactic magnetic field may be affected by large-scale structures such as ``filaments'', ``voids'', and ``clusters''. The convolution analysis of galaxy distributions and magnetic field structure is necessary to connect UHECR arrival directions to sources along the large-scale structure \citep{Hackstein2018}. 

The magnetic structure may affect the multiplet analysis approach. 
\cite{He2016} developed a new technique to constrain the single source and magnetic field structure from the pattern of arrival direction and energies of UHECRs, based on the assumption that the magnetic field structure changes the traces of multiplets.  
Based on this approach, \cite{He2024} reported the multiplet in the public Auger dataset and suggested the Sombrero Galaxy as the source. 
In these cases, the multiplets are not correctly evaluated with a simple circle with the same search radius $\lambda_{0}$. 
Applying the magnetic field structure to the model and analysis method, with deflection considered, may provide additional insight into our work. 

\subsection{Future prospects for observations}
The direct observation of $X_{\rm max}$ by fluorescence detectors (FDs) is important to determine the mass composition of UHECRs. 
Current FD observations in the TA experiment and the Pierre Auger Observatory do not report mass composition above 100 EeV. 
Future reports on $X_{\rm max}$ from TA and TA$\times$4 \citep{TAx4} experiments and AugerPrime \citep{AugerPrime} are expected to constrain the mass composition at higher energies. 
The higher-statistics UHECR dataset above 100 EeV with full-sky exposure is mandatory for precise measurements of small-scale anisotropies.
For further investigation, TA$\times$4 \citep{TAx4} and AugerPrime\citep{AugerPrime}, and The Global Cosmic Ray Observatory \citep[GCOS]{GCOS2022} will take an important role. 

\section{conclusion}
We searched for small-scale anisotropy (doublet) in the public dataset from the Pierre Auger Observatory \citep{Auger2022}. We confirmed there is no significant small-scale anisotropy within $3\tcdegree$ for UHECRs with energies above 100 EeV. This indicates an almost isotropic sky distribution. 

To interpret the non-detection of the anisotropy, we have tested scenarios in which UHECRs are primarily composed of light nuclei (such as protons), heavy nuclei (such as iron), or super-heavy nuclei (elements heavier than iron), using mock UHECR observations from the semi-analytical galaxy catalog and assuming turbulent magnetic fields. 

We set the lower limits on UHECR source density $\rho$ and magnetic field strength (${\rm \alpha} = \left( \frac{\lambda}{1\:\rm Mpc} \right)^{1/2}  \left(\frac{B}{1\:\rm nG} \right)$). 
In the absence of magnetic fields, the isotropy excludes the pure-proton and iron cases. 
The pure-uranium case can explain the current observation with turbulent magnetic fields when the source density is high (ex. $\rho_{\rm source}=10^{-4.5}\: {\rm Mpc}^{-3}$). When the turbulent field is strong, the pure-iron and uranium cases are consistent with the observation. 
Almost all the parameter set $(\rho,\alpha)$ within observational constraints rejects the pure-proton case. 
The realization of the magnetic field structure in the simulation and future reports on $X_{\rm max}$ will play an important role in discriminating between the super-heavy mass composition scenario and other scenarios. 
\begin{acknowledgments}
We thank the fruitful discussions and comments from members of RIKEN and the Institute for Cosmic Ray Research (ICRR). 
We also thank the joint research program of the Institute for Cosmic Ray Research (ICRR), the University of Tokyo.
E.~K. and S.~N. acknowledge the support from the Pioneering Program of RIKEN for Evolution of Matter in the Universe (r-EMU). 
R.H. is supported by JSPS KAKENHI Grant Numbers JP25K17426, JP25H01294. 
R.H. acknowledges the supports from RIKEN Special Postdoctoral Researchers Program (SPDR). 
E.K. is supported by JSPS Grant-in-Aid Scientific Research (KAKENHI) (C), Grant Number JP25K07333, and also supported by the Chinese Academy of Sciences President’s International Fellowship Initiative, Grant No. 2025PVC0021.
The numerical calculations have been partially performed on the HOKUSAI BigWaterfall2 (HBW2) Supercomputer at RIKEN.
S.~N. is supported by the ASPIRE project for top scientists, JST ’RIKEN-Berkeley mathematical quantum science initiative.’
S.N. is supported by JSPS Grant-in-Aid Scientific Research (KAKENHI) (A), Grant Number JP25H00675, and (B), Grant Number JP23K25874.
S.N. is supported by the JST ASPIRE Program “RIKEN-Berkeley mathematical quantum science initiative”.
\end{acknowledgments}


\bibliography{bibfile}
\bibliographystyle{unsrt}

\end{document}